\documentclass{jors}

\usepackage[numbers,round,sort&compress]{natbib}
\usepackage{booktabs}
\usepackage{tabularx}
\newcolumntype{P}[1]{>{\raggedright\arraybackslash}p{#1}}
\newcolumntype{Y}{>{\raggedright\arraybackslash}X}
\usepackage{graphicx}
\usepackage{microtype}

\definecolor{mygray}{gray}{0.6}

\hypersetup{colorlinks=true,linkcolor=black,citecolor=black,urlcolor=blue}

\newcommand{\code}[1]{\texttt{#1}}

\begin{document}

\rule{\textwidth}{1pt}

\section*{(1) Overview}

\section*{Title}

\textbf{vst-bench-ml: Auditable Dataset Generation for Audio Plugins}

\section*{Paper Authors}

Youssef Hemimy

\section*{Paper Author Roles and Affiliations}

Youssef Hemimy: software design and development, validation, documentation, and
manuscript review and editing; St. Thomas University.

\section*{Abstract}

Generating datasets with audio plugins requires preserving how each audio observation
was produced. \code{vst-bench-ml} is a Python package that samples plugin parameters,
coordinates rendering through a separately installed host, and records software
identities, rendering settings, and output hashes for later verification. A sealed
manifest binds this evidence and supports offline integrity checks without loading
the plugin. A companion repository provides a sample-free VST3 test plugin, a
deterministic offline host, and additional checks linking rendering attempts to
recorded outputs. On a pinned macOS arm64 and Python 3.12 stack, functional
validation of the pre-release versions reproduced package and host tests. Two runs in
fresh processes reproduced the same render-plan hash and all eight expected output
hashes, covering four sampled parameter configurations and two fixed MIDI signals;
12 negative cases also passed. Later licensed 0.1.0 release candidates passed
producer-run continuous integration and local validation. Version 0.1.1 adds public
Git source repositories, package-index installation, and user documentation; a
separate release check reproduced the expected plan and output hashes. These checks establish
integrity and repeatability within the tested setup; they do not establish portability,
scientific suitability, or perceptual quality.

\section*{Keywords}

research software; audio plugins; dataset provenance; deterministic validation;
VST3; machine-learning datasets

\section*{Introduction}

Audio plugins can generate controlled audio datasets by varying parameters and MIDI
inputs. To audit these datasets, researchers need a record connecting each output to
the software, plugin settings, and rendering conditions that produced it. Without
that record, it is difficult to determine whether observations were generated under
the intended conditions or whether stored outputs have subsequently changed.

Existing Python software supports offline audio-plugin rendering. DawDreamer
constructs graphs of audio processors, including VST plugins, for offline batch
processing \citep{braun2021dawdreamer}; Pedalboard loads VST3 and Audio Unit
instruments and effects from Python \citep{sobot2021pedalboard}. These tools provide
context for the rendering workflows addressed here; this paper does not compare
their provenance capabilities.

\code{vst-bench-ml} supports auditable dataset generation by sampling plugin
parameters, coordinating rendering through a separately installed host, and linking
generated audio to a record of its production conditions. It records retained
observations separately from dropped and failed attempts and verifies stored
artifacts without loading the plugin. A companion repository provides a controlled
test plugin and host, together with additional checks that connect rendering
attempts to their recorded outputs.

This software paper describes the two-repository architecture, its quality controls,
and the conditions under which the software may be reused. It does not report a
model-training experiment, perceptual study, or scientific comparison. Passing the
integrity checks does not establish that a generated dataset is scientifically
suitable for a downstream task.

\section*{Implementation and Architecture}

\subsection*{Separated package and host}

The Python package selects parameter settings and MIDI inputs, requests audio
renders, and records the resulting outputs and provenance. A separately installed
host loads and runs the plugin. The two communicate over localhost using JSON-RPC
2.0 \citep{jsonrpc2010}. This separation allows the package and host to be versioned
independently, with plugin loading, licensing, binary identity, and platform behavior
handled at the host boundary.

The orchestration software targets Python 3.10 or newer \citep{python2026}. NumPy
provides array operations \citep{harris2020numpy}, SoundFile provides Python audio I/O
\citep{soundfile} over libsndfile \citep{libsndfile}, and \code{pyloudnorm} supplies the locked loudness-normalization
implementation \citep{steinmetz2021pyloudnorm}. The package also performs basic
quality checks, exports records, and verifies stored datasets. It does not distribute
a host, plugin binary, dataset, audio, model, preset library, or product schema.

Before rendering, the client asks the host to load the requested plugin and return
host and plugin identities. Required fields include the host protocol version,
source repository label and revision (Git commit or archive hash), executable hash,
and build configuration, together
with the plugin path, hash, format, name, vendor, and stable identifier. The client
compares the resolved plugin path reported by the host with the resolved requested
path; the render command then compares the parameter manifest's plugin identifier
with the identifier returned by the host. Rendering does not begin if either
comparison fails, required identity fields are missing, the host reports an
already-loaded plugin, or it does not advertise deterministic reset and parameter
readback capabilities. The host-reported binary hashes are recorded as provenance.

The three separately controlled components and their rights boundaries are summarized
in Table~\ref{tab:components}. Rights do not transfer across rows.

\begin{table}[!htbp]
\centering
\small
\begin{tabularx}{\textwidth}{@{}P{0.18\textwidth}P{0.25\textwidth}Y@{}}
\toprule
\textbf{Component} & \textbf{Distributed boundary} & \textbf{Rights and evidence boundary} \\
\midrule
\code{vst-bench-ml} & Python orchestration and offline verification & MIT source; no host, plugin, dataset, audio, model, private research module, or third-party binary. \\
Companion repository & Fixture source, offline host, validation scripts, and expected metadata & MIT source; CC0-1.0 applies only to qualifying sample-free generated fixture outputs. \\
External execution inputs & Separately installed VST3 SDK, lawful plugins, and user-selected data & Not relicensed or redistributed; each use must establish its own compatibility and rights. \\
\bottomrule
\end{tabularx}
\caption{Component and rights boundaries.}
\label{tab:components}
\end{table}

\begin{figure}[!htbp]
\centering
\includegraphics[width=\textwidth]{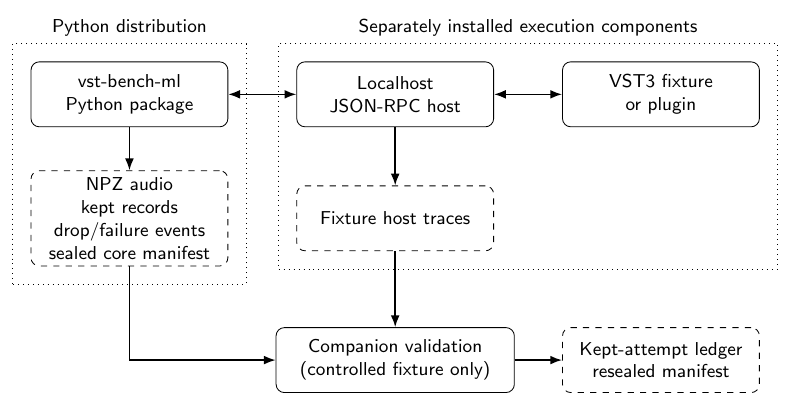}
\caption{System boundary. The package exports NPZ audio, records, events, and a sealed
core manifest. The host runs the plugin and supplies fixture traces; companion
validation adds kept-attempt evidence and reseals the manifest. Source: author's diagram.}
\label{fig:architecture}
\end{figure}

Figure~\ref{fig:architecture} shows the system boundary. The companion host implements
the VST3 lifecycle against the separately retrieved SDK \citep{steinbergvst3}; other
hosts may implement the same protocol, but compatibility is unverified until they
satisfy the identity and behavioral checks.

\textbf{Figure 1 alt text:} Package artifacts and host traces feed companion fixture validation.

\subsection*{Provenance and ledger layers}

The core package writes kept observations to \code{records.jsonl} and appends only
\code{drop} and \code{failure} events to \code{failures.jsonl}. A JSON manifest records
software and environment identities, sampling-plan metadata and its hash, and outcome
counts. It also records the relative path, byte count, and SHA-256 hash of each output
and of the records and failure-ledger files.

Sealing adds a SHA-256 digest of the manifest payload serialized in a canonical JSON
form, excluding the integrity field that stores the digest itself. Offline
verification recomputes this digest and the referenced file hashes, checks byte
counts, and reconciles records and events with the manifest. These checks establish
consistency with the recorded manifest; the seal does not authenticate its origin.
Core verification rejects unknown failure-ledger statuses. Kept observations remain
in the records file, so the core failure ledger alone does not contain a terminal
entry for every kept render.

For the controlled fixture, the companion repository consumes external host traces and
the core kept records, requires an empty drop/failure ledger on the successful path,
and appends one schema-versioned \code{kept} attempt entry per trace. It then binds the
complete attempt ledger into a stricter companion manifest. Its negative-case stage
appends separately identified expected-rejection evidence. This fixture-specific
synthesis and sealing is a companion validation function, not core-package emission.

The schema design follows the general provenance principle that entities, activities,
and their relationships should be represented explicitly \citep{provdm2013}. Table
\ref{tab:provenance} maps the layers to the two implementations.

\begin{table}[!htbp]
\centering
\small
\begin{tabularx}{\textwidth}{@{}P{0.16\textwidth}P{0.37\textwidth}Y@{}}
\toprule
\textbf{Layer} & \textbf{Bound fields} & \textbf{Implementation boundary} \\
\midrule
Execution & Package version and source hashes, Python/runtime identity, dependency versions, normalization method & Core manifest; companion adds Git commit/tree/dirty state or source-archive manifest/file-list hashes, installed-wheel provenance, lock hash, and native-library identities. \\
Host/plugin & Protocol, host source, build and executable identity, reported plugin path and hash, format, name, vendor, identifier & Host-reported identities bound by core; reset/readback capabilities checked at preflight. Companion adds fixture checks. \\
Plan and observation & Parameter-definition hashes, varied indices, seed derivation, plan hash, MIDI signal names, timestamps, kept records, drop/failure events & Core plan metadata, records, event ledger, and sealed manifest. \\
Attempt evidence & Ordered parameter definitions, MIDI hashes, attempt IDs, trace/record correspondence, readback, reset policy, kept disposition, negative cases & Companion-only evidence and complete ledger for the controlled fixture. \\
Artifact integrity & Relative path, byte count, SHA-256, aggregate hashes, count reconciliation & Core offline verification plus stricter companion validation. \\
\bottomrule
\end{tabularx}
\caption{Provenance layers, fields, and implementation ownership.}
\label{tab:provenance}
\end{table}

\begin{figure}[!htbp]
\centering
\includegraphics[width=\textwidth]{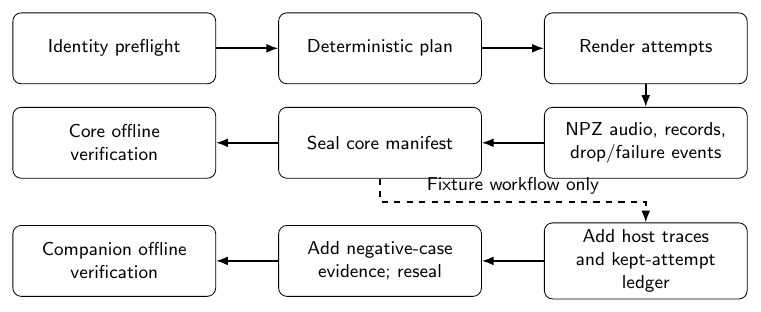}
\caption{Evidence lifecycle. Core artifacts can be sealed and verified without the
companion. The separate fixture workflow combines a successful core run with host
traces, adds kept-attempt and expected-rejection evidence, and reseals for companion
verification. Drop/failure records are preserved by the core; the successful fixture
path requires an initially empty failure ledger. Source: author's diagram.}
\label{fig:lifecycle}
\end{figure}

Figure~\ref{fig:lifecycle} distinguishes standalone core verification from the
additional companion fixture workflow.

\textbf{Figure 2 alt text:} Core verification and companion fixture validation follow separate paths.

\subsection*{Offline verification}

The core \code{verify} command resolves artifact paths beneath the dataset root and
checks the sealed manifest digest, identity completeness, byte counts and SHA-256
hashes, records/output correspondence, outcome consistency, and drop/failure-event
counts. It does not load the plugin or rerender audio. The companion validator
additionally checks complete fixture attempt coverage, host-trace correspondence,
installed-wheel identity, parameter readback, MIDI identity, reset policy, and the
named adversarial cases. These checks establish consistency with the recorded stack,
not scientific validity or resistance to every possible attack.

\section*{Quality Control}

The build and validation stack uses Hatchling \citep{hatchling2026}, pytest
\citep{pytest2026}, Ruff \citep{ruff2026}, CMake \citep{cmake2026}, Git
\citep{git2026}, and AppleClang from the Xcode command-line tools
\citep{xcode2026}.

The project's AI research agent performed a separate internal functional validation pass
using clean clones checked out at the tested pre-release commits.\textsuperscript{\ref{note:validation-versions}}
This pass built the Python wheel and source distribution, installed 19 dependencies
with locked versions and artifact hashes, built the fixture and offline runner, ran
unit tests and lint checks, rendered twice in fresh processes, and exercised the
negative cases summarized in Table~\ref{tab:quality}.

The later licensed 0.1.0 release candidates passed package and host continuous
integration checks and producer-run local validation of the release workflow.
Those checks apply to the later versions and are distinct from the pre-release
validation summarized below.

A separate AI-assisted release validation on 29 September 2026 checked version
0.1.1 from the new public repositories (package commit \code{25c839ba}, companion
commit \code{783e1d9}). Package tests (31) and companion tests (36) passed, as did
wheel/source-distribution metadata checks and installed-wheel validation. Two
fresh render processes reproduced the existing sampling-plan digest and all eight
expected output hashes; the twelve negative cases also passed. The runtime
rendering algorithms and native fixture implementation were unchanged; the
release corrected the module version, public repository identity, and the
validator's source-file boundary for the public checkout. These release checks
are separate from the historical results in Table~\ref{tab:quality}.

\begin{table}[!htbp]
\centering
\small
\begin{tabularx}{\textwidth}{@{}P{0.22\textwidth}Y@{}}
\toprule
\textbf{Gate} & \textbf{Observed result on pinned stack} \\
\midrule
Package tests & 31 passed; wheel/source build and clean wheel installation passed. \\
Host tests & Four passed (pre-release); fixture and offline runner built externally to the repository. \\
Deterministic workflow & Two fresh processes each rendered and kept eight records from four parameter configurations and two MIDI signals; both reproduced the same expected plan hash and all eight expected output-byte hashes. \\
Negative cases & 12 passed, covering substituted package artifacts, wrong plugin identity, unreachable host, malformed parameters, missing normalization, out-of-range values, reset/readback mismatch, duplicate attempts, corrupted output, and incomplete-ledger evidence. \\
Static quality & Configured Ruff lint checks passed with cache disabled. \\
\bottomrule
\end{tabularx}
\caption{Quality-control results from the pre-release validation pass on the pinned macOS arm64 and Python 3.12 stack. Later release-candidate checks were producer-run.}
\label{tab:quality}
\end{table}

The complete plan and output digests are recorded in
\path{fixtures/expected-output.json}, included in the public
\code{vst-bench-dataset-host} archive listed under Software Location.
Note~\ref{note:validation-versions} identifies the historical validation commits.
The sample fixture exposes six parameters:
three floating-point controls, one finite-choice control, one integer control, and one
boolean control. One floating-point control is intentionally audio-irrelevant, while
another has waveform-dependent behavior; these are behaviors rather than additional
parameter types.
Generated validation artifacts remain outside both source repositories.

\section*{(2) Availability}

\section*{Operating System}

The package and companion workflow were validated on macOS with arm64 hardware.
General rendering portability across operating systems and architectures has not
been established.

\section*{Programming Language}

The package metadata declares Python 3.10 or newer as the minimum requirement.
Local release validation used Python 3.12.13; the documented workflow targets
Python 3.12. Support beyond tested environments has not been established. The
companion fixture and runner use C++17, with CMake as the build system.

\section*{Additional System Requirements}

Rendering requires a separately installed host implementing the documented localhost
protocol and a compatible plugin. Offline verification requires the Python package
and stored dataset artifacts, without a host or plugin.

Building the companion fixture and runner requires CMake 3.25 or newer, a C++17
compiler, Git, network access to retrieve the pinned VST3 SDK source, and build
storage outside the source repositories. The tested build used CMake 4.2.3 and
AppleClang 21.0.0 from the Xcode command-line tools. Instructions for creating the
Python environment, building the host, and generating a first dataset are provided
in the public companion's \path{docs/QUICKSTART.md}; its \path{docs/VALIDATION.md}
describes the complete locked validation workflow. The original source ZIPs retain
their historical archive instructions.

\section*{Dependencies}

The package pins its direct runtime dependencies to NumPy 2.2.6, SoundFile 0.13.1,
and \code{pyloudnorm} 0.1.1. Build and validation tools include Hatchling 1.27.0,
pytest 8.4.1, and Ruff 0.12.0. The package's \path{requirements/lock.txt} records
these tools and dependencies, including transitive dependencies, by version and
artifact hash. The companion repository pins the VST3 SDK revision in
\path{CMakeLists.txt} and records its required submodule revisions in
\path{THIRD_PARTY.md}.

\section*{List of Contributors}

Youssef Hemimy (St. Thomas University), sole human contributor: software design and
implementation, validation, documentation, and maintenance.

\section*{Software Location}

Version 0.1.1 is distributed through public Git repositories, with browsable source,
line-level history, release tags, and issue reporting. The Python package is also
available from PyPI. The original 0.1.0 source archives remain available from Zenodo
\citep{hemimy2026vstbenchml,hemimy2026vstbenchhost}; those DOIs identify 0.1.0 only.

\textbf{Current source and package distribution (0.1.1)}

\begin{description}[noitemsep,topsep=0pt]
\item[Package source:] \url{https://github.com/itsuzef/vst-bench-ml-public}
\item[Companion source:]\hfill\\
\url{https://github.com/itsuzef/vst-bench-dataset-host-public}
\item[Package index:] \url{https://pypi.org/project/vst-bench-ml/0.1.1/}
\item[Release tags:] \code{v0.1.1} in both repositories.
\item[Licence:] MIT source; companion output terms remain as stated below.
\end{description}

The public histories begin with imports of the published 0.1.0 source snapshots,
identified by \code{archive-0.1.0} tags and documented archive hashes. Subsequent
changes are tracked as public Git commits; earlier private development history
is not included.

\begin{samepage}
\textbf{Package archive}

\begin{description}[noitemsep,topsep=0pt]
\item[Name:] vst-bench-ml: Auditable Dataset Generation for Audio Plugins
\item[Persistent identifier:] \path{10.5281/zenodo.22924036}
\item[Record and download:] \url{https://zenodo.org/records/22924036}
\item[Licence:] MIT
\item[Publisher:] Zenodo
\item[Version published:] 0.1.0
\item[Date published:] 23 September 2026
\end{description}
\end{samepage}

\begin{samepage}
\textbf{Companion archive}

\begin{description}[noitemsep,topsep=0pt]
\item[Name:] vst-bench-dataset-host: Sample-Free VST3 Fixture and Offline Validation Host
\item[Persistent identifier:] \path{10.5281/zenodo.22924073}
\item[Record and download:] \url{https://zenodo.org/records/22924073}
\item[Licence:] MIT source; CC0-1.0 only for qualifying sample-free generated fixture outputs
\item[Publisher:] Zenodo
\item[Version published:] 0.1.0
\item[Date published:] 23 September 2026
\end{description}
\end{samepage}

The companion archives support source-snapshot verification; current Git checkouts
record Git source identities instead. External SDK source is retrieved separately.
Neither the repositories nor the archives include compiled plugin binaries,
datasets, or generated audio.

\section*{Installation and User Documentation}

In a Python virtual environment, users install the package with
\code{python -m pip install vst-bench-ml==0.1.1}. Pip installs the declared runtime
dependencies. Rendering additionally requires the separately built host and a
compatible plugin; installing the Python package alone supports offline verification
of existing dataset artifacts.

The package's \href{https://github.com/itsuzef/vst-bench-ml-public}{README landing page}
links to the user guide in \path{docs/README.md}, covering installation, commands,
output files, and troubleshooting. The companion's
\href{https://github.com/itsuzef/vst-bench-dataset-host-public/blob/main/docs/QUICKSTART.md}{first-dataset walkthrough}
covers native compilation, starting the local host, rendering the sample-free
fixture, and verifying eight generated outputs. Architecture and protocol references
are provided separately. The ordinary user workflow is distinguished from the
complete release-validation procedure and its pinned environment.

\section*{Language}

English (documentation and command-line messages).

\section*{(3) Reuse Potential}

The validated workflow demonstrates generation of parameterized audio observations
with the controlled fixture, offline verification of core-format datasets, and the
companion's conformance checks on the pinned stack. The manifest and kept records
support later auditing, while the event stream preserves records of dropped or failed
attempts. Offline verification can be used without loading a host or plugin. The
companion additionally demonstrates a complete attempt ledger linked to host traces
for its controlled fixture runs.

Potential extensions include using other plugins, implementing another host, and
adapting quality rules to a research task. These uses require further validation.
A new host must implement the documented identity, reset, readback, and rendering
contract and establish compatibility on its intended platform. Modified quality
rules require explicit thresholds, versioning, tests, and provenance. Each extension
must also address the rights applicable to its plugins, binaries, and generated
audio. Table~\ref{tab:reuse} distinguishes demonstrated capabilities from proposed
extensions and states the evidence limits for each path.

\begin{table}[!htbp]
\centering
\small
\begin{tabularx}{\textwidth}{@{}P{0.20\textwidth}P{0.31\textwidth}Y@{}}
\toprule
\textbf{Reuse path} & \textbf{Requirements} & \textbf{Evidence and limits} \\
\midrule
Generate observations & Lawful plugin, compatible host, parameter manifest, pinned environment & Demonstrated with the controlled fixture; other plugins, hosts, and cross-machine output identity are unverified. \\
Verify a core dataset & Manifest, records, drop/failure events, and referenced artifacts & Demonstrated integrity checks; no complete kept-attempt trace or assessment of scientific fitness or perceptual quality. \\
Run companion conformance & Pinned fixture, host trace, package wheel/lock, expected metadata & Demonstrated on the pinned stack; the fixture does not establish representativeness of commercial plugins or musical data. \\
Implement another host & Protocol identity, reset, readback, and render behavior & Proposed extension; each implementation requires compatibility and behavioral validation. \\
Extend quality rules & Explicit thresholds, versions, tests, and provenance & Proposed extension; each rule set requires validation for its intended research task. \\
\bottomrule
\end{tabularx}
\caption{Reuse paths, requirements, and evidence distinguishing demonstrated capabilities from proposed extensions.}
\label{tab:reuse}
\end{table}

The public source repositories accept questions and bug reports through Issues,
and source or documentation improvements through pull requests. Each repository
includes contribution instructions. The earlier
\href{https://github.com/itsuzef/vst-bench-public}{public resources page} retains
0.1.0 archive checksums and links to current sources and documentation.
No response-time or ongoing-support guarantee is made.

\section*{Data Accessibility}

The companion source repository and archive include the fixture parameter definitions, fixed MIDI
input definitions, expected plan and output hashes, and instructions for regenerating
the validation outputs on the documented stack. Generated audio and validation-run
artifacts are not included in the public archives. Qualifying sample-free generated
fixture outputs are covered by the companion's \path{OUTPUT-LICENSE}; this does not
extend to third-party plugins or audio.

\section*{Limitations}

The historical internal validation covered one macOS arm64 and Python 3.12.13 stack, one
self-authored fixture, four sampled parameter configurations, and two fixed MIDI
signals, producing eight outputs per run. Two runs in fresh processes reproduced
the expected sampling-plan hash and output hashes on that setup. This demonstrates
repeatability under the tested conditions. The later release checks are also
scoped to their recorded environments; general reproducibility across machines,
platforms, or alternative hosts and plugins has not been established.

The validation did not measure runtime performance, usability, external adoption,
perceptual quality, or downstream model performance. Integrity checks were exercised
against the enumerated negative cases; passing those cases does not establish
coverage of every possible failure or provide a security guarantee. The fixture's
rights boundary does not transfer to third-party plugins or audio. Each research
use must document its own scientific design, data rights, and compatibility
evidence.

\section*{Funding Statement}

This work was self-funded by the author. No external funding was received.

\section*{Competing Interests}

The author declares no competing interests.

\section*{Author Contributions}

Youssef Hemimy: software design and development, validation, documentation, and
manuscript review and editing.

\section*{Use of Generative Artificial Intelligence}

Claude and Codex were used as assistants for manuscript drafting and software
implementation. An AI research agent also assisted with the internal validation
workflow described in the Quality Control section. The author reviewed and edited
all AI-assisted work and takes responsibility for the software and manuscript.

\section*{Notes}

\begin{enumerate}[noitemsep,topsep=0pt]
\item \label{note:validation-versions} The pre-release validation used package commit
\code{129c0d1} and companion commit \code{1346e2ce}. Later 0.1.0 release-candidate
checks used package commit \code{02b281a9} and companion commit \code{27d03a4e}.
The published companion archive includes source-snapshot verification added after
those commits; these commit identifiers describe the earlier validation stages,
not the complete archived companion source.
The companion workflow pins VST3 SDK commit
\path{3cdf9ca5d1f5b1b21e0a86832aa4abe55607bd96}.
\end{enumerate}

\end{document}